\documentclass{XrU2005}
\usepackage{pstricks,amsmath,graphicx}
\usepackage[varg]{txfonts}

\def\1e{\mbox{1E\,0657--56}}
\def\kms        {km$\;$s$^{-1}$}
\def\chandra    {\emph{Chandra}}
\def\as         {$^{\prime\prime}$}
\def\lax{\lesssim}

\def\bi{\bfseries\itshape}
\def\msun       {$M_{\odot}$}

\title{\chandra\ observation of the most interesting cluster in the
  Universe}

\author{M. Markevitch}

\affil{Harvard-Smithsonian Center for Astrophysics}

\begin{document}


\maketitle

\begin{abstract}

\chandra\ has recently observed \1e, a hot merging system at $z=0.3$ (the
``bullet'' cluster), for 500 ks. I present some of the findings from this
dataset.  The cluster exhibits a prominent bow shock with $M=3.0\pm0.4$ (one
of only two known $M\gg 1$ shock fronts), which we use for a first test of
the electron-ion equilibrium in an intergalactic plasma.  The temperatures
across the shock are consistent with instant shock-heating of the electrons;
at 95\% confidence, the equilibration timescale is much shorter than the
collisional Spitzer value. Global properties of \1e\ are also remarkable.
Despite being extremely unrelaxed, the cluster fits well on the $L_X-T$
relation, yet its total mass estimated from the $M-T$ relation is more than
twice the value measured from lensing.  This is consistent with simulations
predicting that in the middle of a merger, global temperature and X-ray
luminosity may be temporarily boosted by a large factor.

\end{abstract}

\section{Introduction}

\1e\ is the most interesting cluster in the Universe, as officially
confirmed by \chandra\ Peer Review (Anonymous 2003). This system, located at
$z=0.3$, has the highest X-ray luminosity and temperature and the most
luminous radio halo of all known clusters. It is also a spectacular merger
occurring almost exactly in the plane of the sky (Markevitch et al.\ 2002),
and contains one of only two known cluster shock fronts with a Mach number
substantially greater than 1 (the other one is A520 with $M=2$). \chandra\ 
has recently observed it with a 500 ks total exposure. The image from that
dataset is shown in Fig.\ 1. It shows a prominent bow shock preceding a
small, cool ``bullet'' subcluster flying west after passing through a core
of a bigger cluster and disrupting it. In this paper, two interesting
(preliminary) findings from this new dataset are presented, one regarding
the global properties of the cluster, and another based on the
high-resolution electron temperature profile across the shock front. All
errors are 68\%; the assumed cosmology is $h=0.7$, $\Omega_0=0.3$ and
$\Omega_\Lambda=0.7$.

\begin{figure*}[t]
\pspicture(0,0)(17,17)
\rput[tl]{0}(0,17){%
\includegraphics[width=1.0\linewidth,bb=0 14 737 751,clip]%
{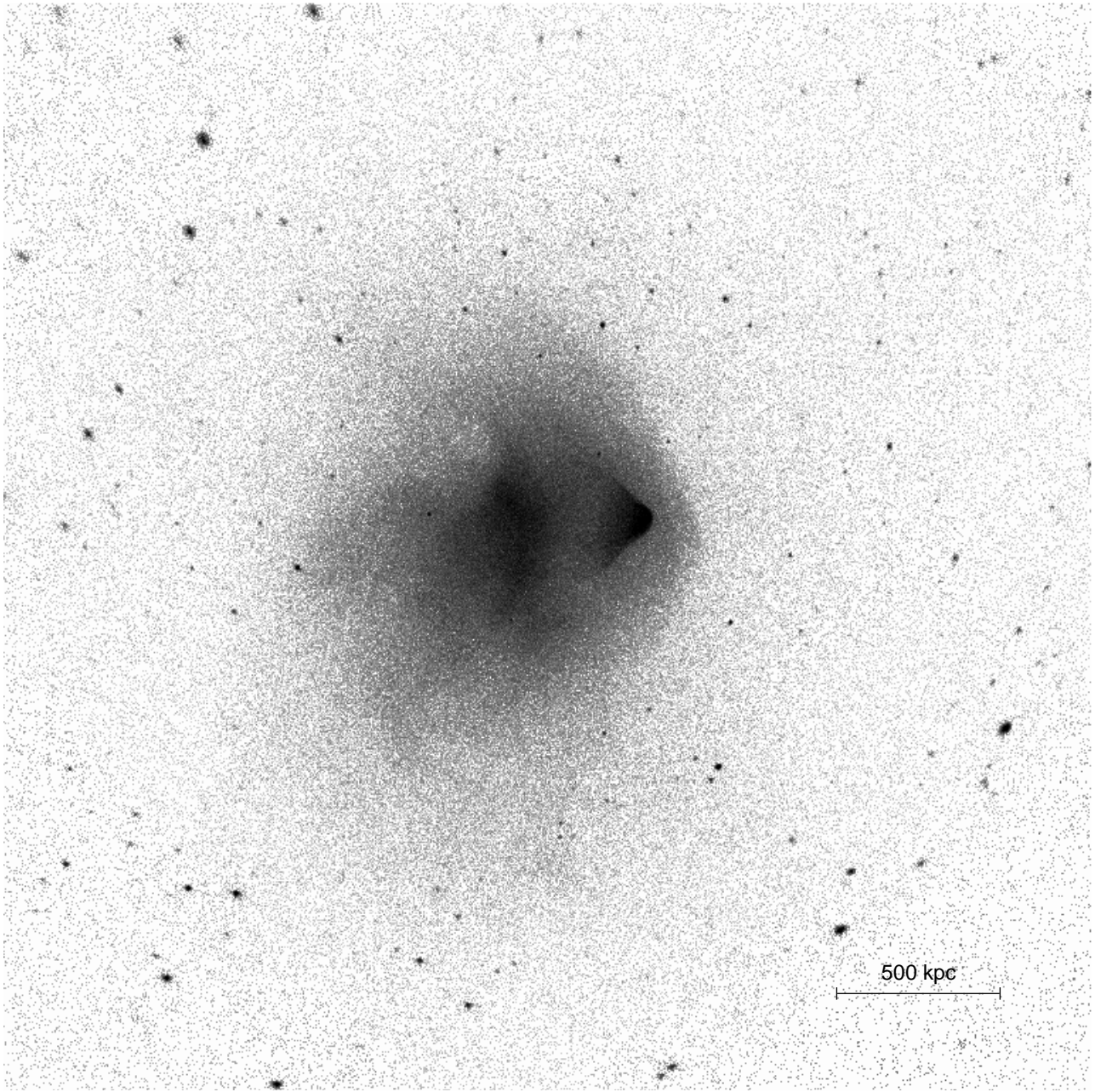}%
}

\psline[linewidth=0.5pt,tbarsize=5pt 1]{|-|}(13.0,1.5)(15.57,1.5)
\rput[tc]{0}(14.29,1.8){\sf 500 kpc}

\endpspicture

\caption{500 ks Chandra ACIS-I image of \1e\ in the 0.8--4 keV band.}
\end{figure*} 

\begin{figure}[b]
\centering
\includegraphics[width=1.0\linewidth,bb=35 178 550 640,clip]%
{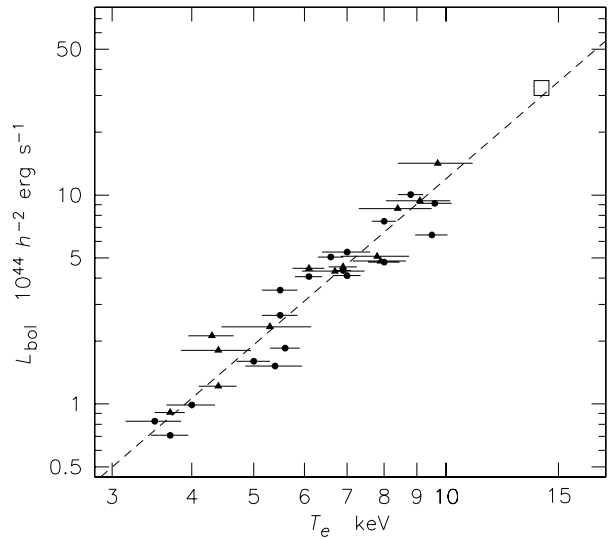}
\caption{\1e, shown as open square, overlaid on the local $L_X-T$ relation
  (with its best-fit power law shown by dashed line), after a correction of
  $L_X$ of this cluster for the redshift evolution of the relation.}
\end{figure} 

\section{An overheated cluster}

A single-temperature fit to the overall ACIS spectrum for the cluster
(excluding a small region around the bullet for consistency with the
``cooling flow corrected'' temperatures for nearby clusters) is
$T=14.1\pm0.2$ keV.  As shown in Fig.\ 2, it fits perfectly on the $L_X-T$
relation for local clusters (Markevitch 1998) after a correction for its
redshift evolution (Vikhlinin et al.\ 2002). However, its total mass
estimated from an X-ray $M_{500}-T$ relation (Vikhlinin et al.\ 2005b; Kotov
\& Vikhlinin 2005), $M_{500}=1.9\times 10^{15}$\msun, is a factor of 2.4
higher than the value within the same radius estimated from weak lensing
(Clowe et al.\ 2004). Given the ongoing violent merger, this is not
unexpected --- simulations have predicted that during a major merger, the
cluster may experience a transient boost of temperature by a factor of
several, lasting of order 0.1~Gyr around the moment of the subcluster core
passage (Randall et al.\ 2002; Rowley et al.\ 2004). In the course of this
rapid change, the cluster moves approximately along the $L-T$ relation.  We
know from the shock velocity (Markevitch et al.\ 2002) that the core passage
in \1e\ has indeed occurred about 0.15~Gyr ago, so this cluster appears to
illustrate precisely this short-lived phenomenon.

\begin{figure}
\centering
\includegraphics[width=1.0\linewidth,bb=40 180 555 692,clip]%
{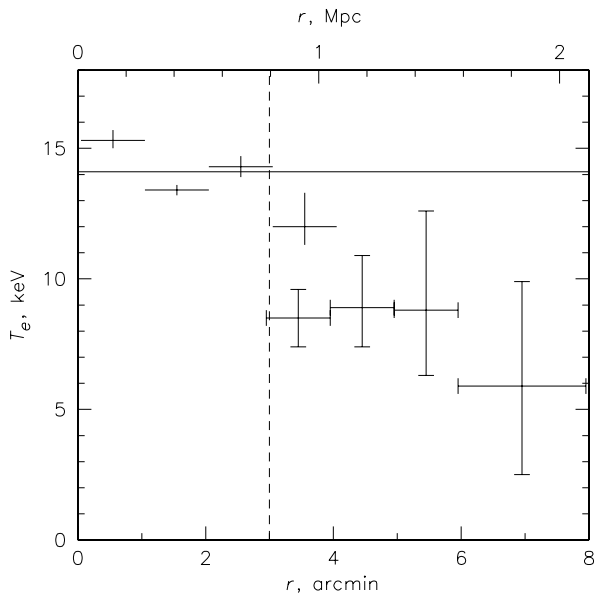}
\caption{Radial temperature profiles for the disturbed region behind the
  shock (crosses without bars) and for the undisturbed region in front of
  the shock (barred crosses). Both profiles are centered on the main
  subcluster. The vertical dashed line approximately shows the shock front;
  the horizontal line is the average temperature.}
\end{figure} 

It is interesting to try and recover a ``pre-merger'' temperature of the
main subcluster. The gas before (west of) the shock front should not yet
know anything about the merger, and stay undisturbed in the gravitational
potential of the main cluster. We have therefore divided the cluster into
the post-shock and pre-shock regions (approximately along the bow shock) and
extracted radial temperature profiles in these regions, both centered at the
mass centroid of the main subcluster (Clowe et al.\ 2004).  The resulting
profiles are shown in Fig.\ 3. If before the merger, the main cluster had a
declining radial profile similar to that in most clusters (Vikhlinin et al.\ 
2005a), extrapolating the ``pre-shock'' profile inwards would give an
average temperature around 10 keV or less, compared to the present 14 keV.
For such a temperature, $M_{500}$ from the $M-T$ relation is within a factor
of 1.5 of the lensing mass, which is within the uncertainty of the Clowe et
al.\ (2004) lensing measurement. (The lensing mass would imply $T\simeq 8$
keV.)

The ``overheated'' cluster \1e\ may thus serve as a cautionary example for
projects involving X-ray surveys of distant clusters, where total masses are
proposed to be derived from low-statistics X-ray data using temperature
profiles, average temperatures or even fluxes. If \1e\ was placed at $z=1$
and observed with \chandra\ with a relatively long 100 ks exposure, its
extremely unrelaxed state would be difficult to detect from the X-ray data
(image, $L_X-T$ relation, etc.), especially if the merger was not oriented
so fortunately in the plane of the sky.  Its mass estimate would of course
be significantly wrong.

\section{Electron-ion equilibrium}

The bow shock in \1e\ offers a unique experimental setup to determine
whether electrons in the intracluster plasma are directly heated by shocks,
or compressed adiabatically and then heated to an equilibrium temperature
via collisions with protons (that are heated dissipatively by the shock).
The collisional equilibration occurs on a Spitzer timescale (e.g., Zeldovich
\& Raizer 1966)
\begin{equation}
\tau_{\rm ep}= 2\times 10^8\;{\rm yr}
\left(\frac{n_e}{10^{-3}\;{\rm cm}^{-3}}\right)^{-1}
\left(\frac{T_e}{10^8\,{\rm K}}\right)^{3/2}
\end{equation}
(note that, conservatively for our measurement, this is 3 times shorter than
the formula given in Sarazin 1988). We cannot measure $T_i$ in X-rays, only
$T_e$.  However, because the shock in \1e\ propagates in the plane of the
sky, we can accurately measure the gas density jump across the front and use
it to predict the post-shock adiabatic and shock-heated (dissipative)
electron temperatures from the pre-shock temperature (using the adiabat and
the Rankine-Hugoniot jump conditions, respectively), and compare it with the
observation. Furthermore, we also know the downstream velocity of the
shocked gas flowing away from the shock. This flow effectively unrolls the
time dependence of the electron temperature along the spatial coordinate for
us.  The Mach number of the shock is conveniently high, so that the
adiabatic and dissipative electron temperatures are sufficiently different
for us to distinguish (e.g., for $M\lax 2$, they would be almost the same).
It is also not a strong shock, for which the density jump would just be a
factor of 4 (for ideal monoatomic gas) and would not let us directly
determine $M$.  Furthermore, the distance traveled by the post-shock gas
during the time given by eq.\ 1, $\Delta x\simeq 230$ kpc $=50''$, is
well-resolved by \chandra.  The statistical quality of the 500 ks dataset is
just sufficient --- in fact, this test was the main science driver for this
long observation.

\begin{figure}
\centering
\includegraphics[width=1.0\linewidth,bb=30 186 533 680,clip]%
{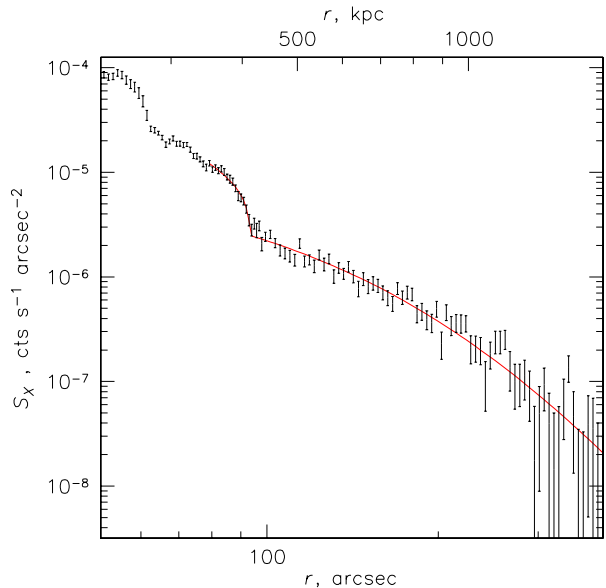}
\caption{X-ray brightness profile  across the shock
  front.  The line shows the best-fit model (a projected sharp spherical
  density discontinuity at the shock).}
\end{figure} 

\begin{figure*}[t]
\hspace{-2mm}
\includegraphics[width=0.48\linewidth,bb=42 186 548 685,clip]%
{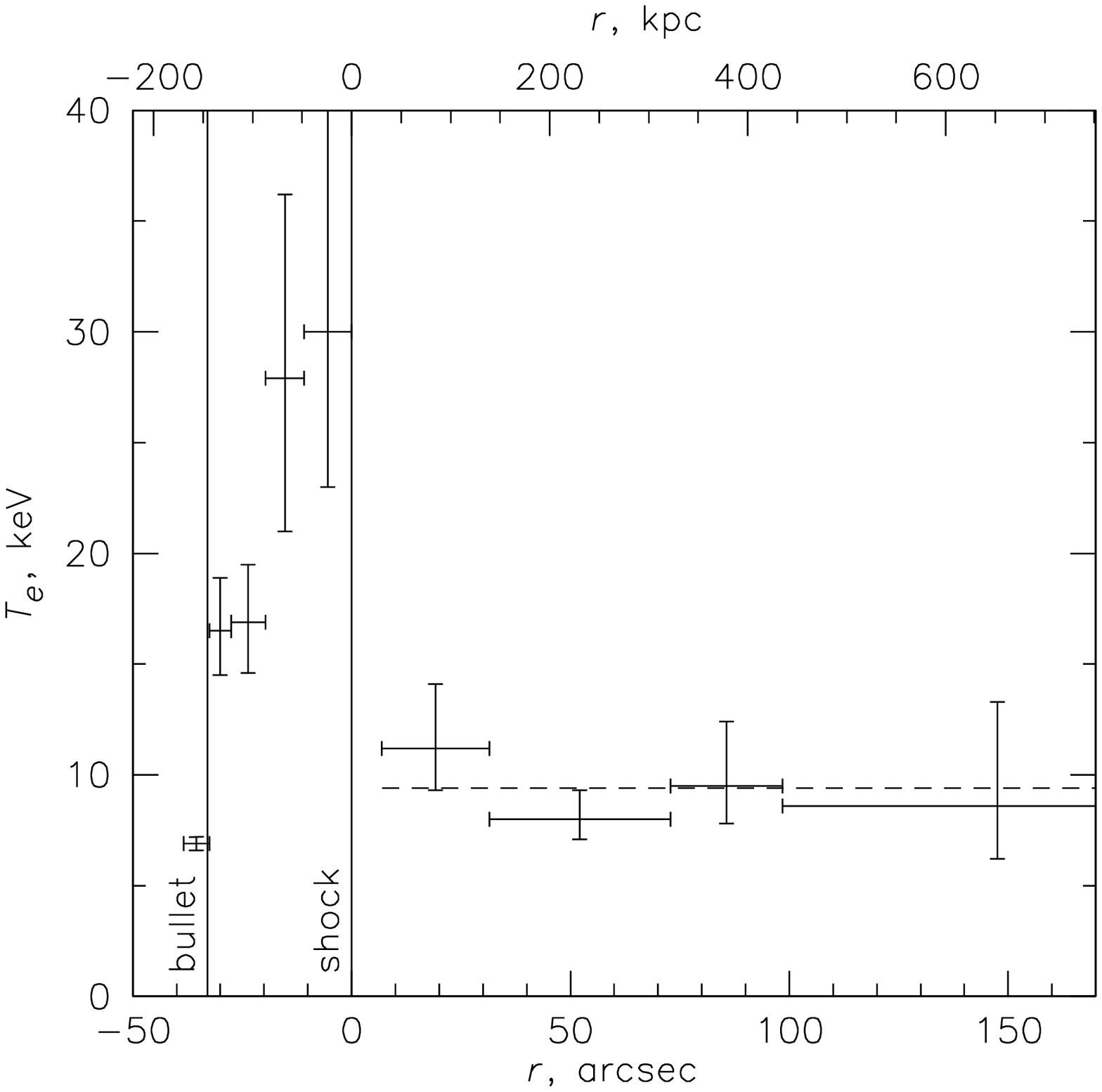}
\hspace{5mm}
\includegraphics[width=0.48\linewidth,bb=42 186 548 685,clip]%
{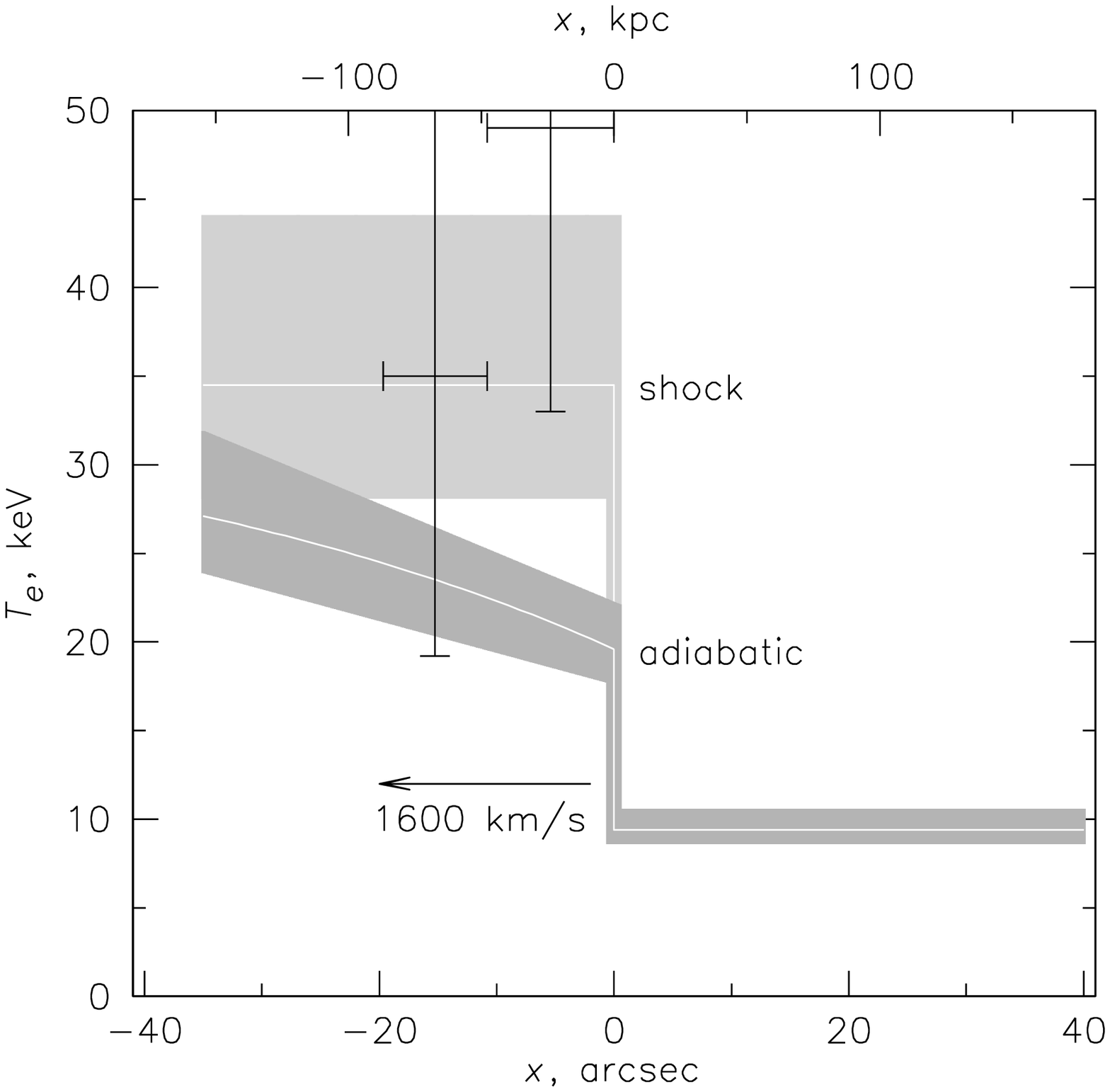}

\caption{Left: {\bi projected} temperature profile in a narrow sector across
  the shock. Two crosses in the shock region with lower temperatures
  correspond to an apparent additional edge-like structure and are not used.
  Vertical lines show the boundaries of the cool bullet and the shock;
  dashed line shows the average pre-shock temperature. Right: {\bi
    deprojected} temperatures for the two post-shock bins overlaid on the
  model predictions (with error bands) for shock-heated and adiabatic
  electron temperatures.  The velocity shown is for the post-shock gas
  relative to the shock.}
\end{figure*} 

Fig.\ 4 shows a 0.8--4 keV surface brightness profile in a narrow sector
across the shock, centered on the center of curvature of the front.  The
inner bump is the bullet (its boundary is a ``cold front''). The edge at
90\as\ is the shock front.  There is also a subtle secondary edge between
these main features, which may also be seen in the deep image. It is
unrelated to the shock, so we should take care to exclude it from any fits
aimed at deprojecting the shock temperatures and densities. The line in
Fig.\ 4 shows a best-fit model consisting of the projected abrupt spherical
density jump at the shock (by a factor of 3.0), a power-law profile inside
the shock and a beta-model outside.  The fit is perfect (the inclusion of
the secondary edge does not affect it).  From this density jump, we
determine $M=3.0\pm0.4$, which corresponds to a shock (and bullet) velocity
of 4700~\kms.

In Fig.\ 5 (left), we show a projected temperature profile in the same
sector across the shock. There is a clear jump of the electron temperature.
At the secondary edge mentioned above, the temperature goes down, which
probably indicates residual cooler gas from the subcluster located ahead of
the bullet.  Therefore, we can only use the two bins closest to the front.
Because we know the gas density profile across the shock, we can accurately
subtract the contributions of the cooler pre-shock gas projected into the
post-shock bins, assuming spherical symmetry.  The large brightness contrast
at the shock helps make this subtraction robust.

The deprojected values are shown in Fig.\ 5 (right). They are overlaid on
the two models (gray bands), one assuming instant equilibration (i.e.,
electrons are heated at the shock), and another assuming adiabatic
compression and subsequent equilibration with protons on a timescale given
by eq.\ 1. (The plot assumes a constant post-shock gas velocity, which of
course is not correct, but we are interested only in the immediate shock
vicinity.)  The deprojected gas temperatures are so high for \chandra\ that
only their lower limits are meaningful. The temperatures are consistent with
instant heating; the ``adiabatic'' model with the Spitzer timescale is
excluded at a 95\% confidence.

A few sanity checks have been performed. The high temperatures are not an
artifact of the deprojection, because the projected temperatures in those
two bins are already higher than the models. We also considered the
possibility of a non-thermal contamination.  The cluster is known to possess
a radio halo (Liang et al.\ 2000) which has an edge right at the shock front
(Markevitch et al.\ 2002). Therefore, there may be an inverse Compton
contribution from relativistic electrons accelerated at the shock.  However,
the power-law spectrum of such emission for any $M$\/ would be {\em
  softer}\/ than thermal, so should not bias our measurements high. We also
extracted a temperature profile in another sector of the shock (away from
the nose), where $M\simeq 2$, and made a similar comparison. The recovered
post-shock temperatures are lower than those in Fig.\ 5 and again in
agreement with the model predictions (either of the two models; for such
$M$, the difference between them is small).  Thus, albeit at a relatively
low significance (95\%), we conclude that the electron-ion equilibration
should be much faster than collisional.

It is of course unfortunate that a cluster with such a perfect geometric
setup and Mach number is so hot that \chandra\ can barely measure the
post-shock temperatures. However, there is no choice of shock fronts and
improvements upon the above measurement are unlikely in the near future.

There hasn't been a measurement of the electron-ion equilibration timescale
in the intergalactic medium before. In the solar wind plasma, the
equilibration is believed to be fast compared to the collisional timescale.
For supernova remnants, which have strong shocks, conclusions vary between
different objects (e.g., Rakowski 2005). Plasma interactions have been
suggested as a fast equilibration mechanism for the solar wind, and it
probably applies for the intracluster plasma as well.

\section{Summary}

From the extra-long 500 ks \chandra\ observation of \1e, we found that this
cluster is observed at a very special, short-lived stage, when its
temperature and luminosity are temporarily boosted by the merger by
significant factors. The total cluster mass estimate from the X-ray $M-T$\/
relation turns out to be more than two times higher than the (presumably)
true mass determined by lensing, indicating a very strong deviation from
hydrostatic equilibrium. If this cluster were at high $z$ and not so well
exposed, it would be difficult to detect its disturbed state, thus it is a
cautionary example for future high-$z$ surveys.

The temperature profile across the shock offers a first test of the
electron-ion equilibrium in an intracluster plasma. The temperatures
indicate that electrons are indeed quickly heated at the shock. The slow
collisional (Spitzer) equilibration rate is excluded at the 95\% confidence.

This unique cluster, well-exposed by \chandra, is the subject of several
other ongoing studies, the results of which will be reported soon.




\begin{thebibliography}{}
 
\bibitem[{{Kakka}(2003)}]{} Clowe, D., Gonzalez, A., \& Markevitch, M.\ 
2004, \apj, 604, 596

\bibitem[{{Kakka}(2003)}]{} Kotov, O., \& Vikhlinin, A. 2005, astro-ph/0511044
 
\bibitem[{{Kakka}(2003)}]{} Liang, H., Hunstead, R. W., Birkinshaw, M., \&
Andreani, P.  2000, ApJ, 544, 686

\bibitem[{{Kakka}(2003)}]{} Markevitch, M.\ 1998, \apj, 504, 27 

\bibitem[{{Kakka}(2003)}]{} Markevitch, M., Gonzalez, A.~H., David, L.,
Vikhlinin, A., Murray, S., Forman, W., Jones, C., \& Tucker, W.\ 2002,
\apjl, 567, L27

\bibitem[{{Kakka}(2003)}]{} Rakowski, C.~E.\ 2005, Adv.\ Space Research, 35,
1017  
 
\bibitem[{{Kakka}(2003)}]{} Randall, S.~W., Sarazin, C.~L., \& Ricker,
P.~M.\ 2002, \apj, 577, 579

\bibitem[{{Kakka}(2003)}]{} Rowley, D.~R., Thomas, P.~A., \& Kay, S.~T.\ 
2004, \mnras, 352, 508

\bibitem[{{Kakka}(2003)}]{} Sarazin, C. L. 1988, X-ray emission from clusters
of galaxies (Cambridge University Press)

\bibitem[{{Kakka}(2003)}]{} Vikhlinin, A., VanSpeybroeck, L., Markevitch,
M., Forman, W.~R., \& Grego, L.\ 2002, \apjl, 578, L107

\bibitem[{{Kakka}(2003)}]{} Vikhlinin, A., Markevitch, M., Murray, S.~S.,
Jones, C., Forman, W., \& VanSpeybroeck, L.\ 2005a, \apj, 628, 655

\bibitem[{{Kakka}(2003)}]{} Vikhlinin, A., Kravtsov, A., Forman, W., Jones,
C., Markevitch, M., Murray, S.~S., \& Van Speybroeck, L.\ 2005b,
astro-ph/0507092

\bibitem[{{Kakka}(2003)}]{} Zeldovich, Y.~B., \& Raizer, Y.~P.\ 1967,
Physics of Shock Waves (New York: Academic Press)
 
\end{thebibliography}
\end{document}